\documentclass[aps,prb,twocolumn,unsortedaddress,superscriptaddress,showpacs]{revtex4-1}
\usepackage{color}
\usepackage{graphicx}
\usepackage{amsmath,amsfonts,amsthm} % Math packages
\usepackage{multirow}

\begin{document}

% Use the \preprint command to place your local institutional report
% number in the upper righthand corner of the title page in preprint mode.
% Multiple \preprint commands are allowed.
% Use the 'preprintnumbers' class option to override journal defaults
% to display numbers if necessary
%\preprint{}

%Title of paper
\title{Structural relaxation and low energy properties of Twisted Bilayer Graphene}
\author{Giovanni Cantele}
\email{giovanni.cantele@spin.cnr.it}
\affiliation{CNR-SPIN, c/o Complesso di Monte S. Angelo, via Cinthia - 80126 - Napoli, Italy}

\author{Dario Alf\`e}
\affiliation{
Dept. of Earth Sciences  and 
London Centre for Nanotechnology 
University College London, 
Gower Street, London, WC1E 6BT, UK}
\affiliation{Universit\`a degli Studi di Napoli ``Federico II'',
Dipartimento di Fisica "Ettore Pancini", Complesso di Monte S. Angelo, via Cinthia - 80126 - Napoli, Italy}
\author{Felice Conte}
\author{Vittorio Cataudella}
\author{Domenico Ninno}
\affiliation{Universit\`a degli Studi di Napoli ``Federico II'',
Dipartimento di Fisica "Ettore Pancini", Complesso di Monte S. Angelo, via Cinthia - 80126 - Napoli, Italy}
\affiliation{CNR-SPIN, c/o Complesso di Monte S. Angelo, via Cinthia - 80126 - Napoli, Italy}
\author{Procolo Lucignano}
\email{procolo.lucignano@unina.it}
\affiliation{Universit\`a degli Studi di Napoli ``Federico II'',
	Dipartimento di Fisica "Ettore Pancini", Complesso di Monte S. Angelo, via Cinthia - 80126 - Napoli, Italy}
%\affiliation{CNR-SPIN, c/o Complesso di Monte S. Angelo, via Cinthia - 80126 - Napoli, Italy}

% repeat the \author .. \affiliation  etc. as needed
% \email, \thanks, \homepage, \altaffiliation all apply to the current
% author. Explanatory text should go in the []'s, actual e-mail
% address or url should go in the {}'s for \email and \homepage.
% Please use the appropriate macro foreach each type of information

% \affiliation command applies to all authors since the last
% \affiliation command. The \affiliation command should follow the
% other information
% \affiliation can be followed by \email, \homepage, \thanks as well.
%\author{}
%\email[]{Your e-mail address}
%\homepage[]{Your web page}
%\thanks{}
%\altaffiliation{}
%\affiliation{}

%Collaboration name if desired (requires use of superscriptaddress
%option in \documentclass). \noaffiliation is required (may also be
%used with the \author command).
%\collaboration can be followed by \email, \homepage, \thanks as well.
%\collaboration{}
%\noaffiliation

\date{\today}

\begin{abstract}
The structural and electronic properties of twisted bilayer graphene are investigated from first principles and tight
binding approach as a function of the twist angle (ranging from the first ``magic'' angle $\theta=1.08^\circ$ to $\theta=3.89^\circ$,
with the former corresponding to the largest unit cell, comprising 11164 carbon atoms).
By properly taking into account the long-range van der Waals interaction, we provide the patterns for the atomic displacements
(with respect to the ideal twisted bilayer). The out-of-plane relaxation shows an oscillating (``buckling'') behavior, 
very evident for the smallest angles, with the atoms around
the AA stacking regions interested by the largest displacements. The out-of-plane displacements are accompanied by a significant
in-plane relaxation, showing a vortex-like pattern, where the vorticity (intended as curl of the displacement field) is reverted
when moving from the top to the bottom plane and viceversa. Overall, the atomic relaxation results in the shrinking of the 
AA stacking regions in favor of the more energetically favorable AB/BA stacking domains.

The measured flat bands emerging at the first magic angle can be accurately described only if the atomic relaxations are taken into account. Quite importantly, the experimental gaps separating the flat band manifold from the higher and lower energy bands cannot be reproduced if only in-plane or
only out-of-plane relaxations are considered.
The stability of the relaxed bilayer at the first magic angle is estimated to be of the order of 0.5-0.9 meV per atom (or 7-10 K).
Our calculations shed light on the importance of an accurate description of the vdW interaction and of the resulting atomic relaxation to envisage the electronic structure of this really peculiar kind of vdW bilayers.
\end{abstract}

% insert suggested PACS numbers in braces on next line
% insert suggested keywords - APS authors don't need to do this
%\keywords{}

%\maketitle must follow title, authors, abstract, \pacs, and \keywords
\maketitle

\section{Introduction}
\label{sec:intro}
After the first experimental findings~\cite{Cao:2018kn,Cao:2018ff},  Twisted Bilayer Graphene (TBG) has been subject of intense investigation from both the  experimental and theoretical point of view.

When the rotation angle between the two graphene layers is close to the first "magic angle" $\theta\sim 1.08^\circ$, transport experiments show  different superconducting domes as well as correlated insulating phases~\cite{Codecido:2019wy,Efetov:2019,Sharpe605,Yankowitz1059,Kennes2020,Choi:2019aa}.

Most of the unconventional transport properties of TBG originate from the  almost flat bands (FBs) at the Fermi energy, originally predicted in Ref.~\cite{Bistritzer:2011ho},
whose  bandwidth, of the order of $\sim 10$ meV, has been confirmed  also from tunnel spectroscopy experiments \cite{Kerelsky:2019aa,Xie:2019aa,Jiang:2019aa,Choi:2019aa}.
The FBs manifold, which can host up to four electrons above the Fermi energy and four holes below it, is  separated by an energy gap of $\sim 50$  meV from both higher and lower energy bands,  and has been clearly observed in  recent nano-ARPES measurements~\cite{Lisi2020direct}. 
When an external gate tunes the system chemical potential within these gaps, a clear band insulating phase appears.
A second, unexpected, insulating phase shows up at half-filling of the FB manifold, both on the electron and on the hole side ($\pm 2$ electrons with respect to charge neutrality). 
The correlated insulating phase is attributed to enhanced electron-electron  interaction within the FBs respectively~\cite{Sboychakov:2018tp,Rademaker:PRB2019}, although some authors are highlighting the relevance of the electron-phonon interaction~\cite{Choi:2018cu,Angeli:2019,Koshino_Son:2019,Lamparski_2020}. 
% maybe driven by electron correlations.
%The origin of this phase remains unclear, it could be likely driven by electron correlations becoming more and  more relevant when bands flatten close to the magic angle, however also other scenarios have been proposed.
After electrostatic doping, achieved by gating the structure, unconventional superconductivity, with a critical temperature 
ranging from 1.7 to 3 K appears in a strong pairing regime, with a phase diagram very similar to that of the underdoped cuprates, whose origin is still to be understood \cite{Angeli:2019,Talantsev:SciRep2020}.

Similar physics is being addressed also in twisted bilayers made out of transition metal dichalcogenides \cite{Maity:PRR_2020}, germanium selenide \cite{Lamparski_2020} other  two dimensional materials ~\cite{PhysRevB.99.155429,PhysRevLett.121.266401}.

That  reveals how the twist angle can be used as a further degree of freedom~\cite{RibeiroPalau:2018ki} 
for combining two-dimensional (2D) materials to implement desired 
properties~\cite{Geim:2014hf, PhysRevMaterials.1.014002,Borriello:2012ja,Cantele:2009de}.
The twisted lattice geometry gives rise to topological
properties of TBG~\cite{Song:2018ul,Hejazi:2019dz,Liu:2018}, unlike 
conventional topological materials~\cite{RevModPhys.82.3045}, where topological properties are mostly due to  spin-orbit interactions~\cite{0034-4885-78-10-106001,PhysRevB.78.035336} and Brillouin zone topology.

In this paper we apply large-scale density functional calculations to better elucidate the origin of the FBs in the single-particle
band structure and show the fundamental role played by the atomic relaxation. 
Relaxation mechanisms have been recently addressed using semiclassical techniques \cite{Guinea:PRB2019,Angeli:2018wy}, 
by contrast in this manuscript we resort to a DFT approach already presented in Ref.\cite{NoiPRB:2019} and find new relaxation patters both at the magic angle and in other low angle twisted structures.
The properties of the FBs manifold, at the first magic angle
is connected  with the atomic displacements originating from the interlayer van der Waals interaction. We show that the
energy gain, induced by the relaxation, becomes of the order of 10 K, much larger than the typical temperature at which 
unconventional superconductivity or the correlated insulating phase are observed in TBG ( $T<$1$\div$2 K). The smaller is the twist angle,
the more pronounced are the atomic displacements with respect to the flat bilayer. In particular, we single out an
oscillating displacement pattern of the out-of-plane displacements at smaller angles, that is smoothed at the larger angles, and a vortex-like in-plane displacement pattern, where the atoms ``rotate" in opposite directions in the two planes.

Tight binding calculations  both at the relaxed and unrelaxed positions are also carried out, to provide a further and less expensive tool to reproduce, especially at the first magic angle, the electronic structure.
We also give the effective parameters that best approximate the ab-initio band structure within the  low energy continuum theory \cite{Bistritzer:2011ho} generalized in the presence of atomic relaxation \cite{Koshino:2018hm}. Interestingly enough, they are largely independent of the twist angle, which makes the continuum model an excellent tool to describe the low energy physics at 
small twist angles. 

The paper is organized as follows: in Sec.~\ref{sec:methods} we outline the technical details of the calculations. In
Sec.~\ref{sec:geometry} we extensively discuss the results on the geometrical relaxation and displacement patterns. 
In Sec.~\ref{sec:electron} the tight binding approach and the continuum model outcomes are compared with the ab initio band structure.
Finally, in Sec.~\ref{sec:conclusions} we summarize our findings and draw our  final conclusions.

\begin{figure*}
    \centering
     %%%% (31,30) - 1.08 degrees
    (a)
    \includegraphics[scale=0.30]{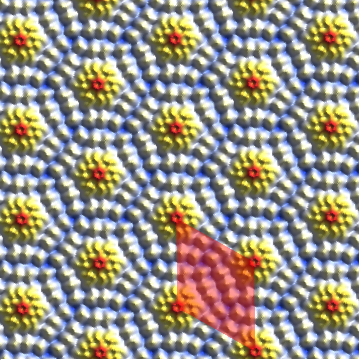}
    \includegraphics[scale=0.50]{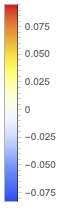}
    \hspace{0.5cm}
    \includegraphics[scale=0.30]{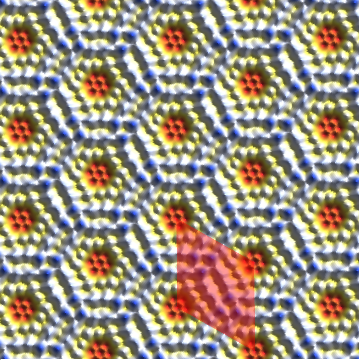}
    \includegraphics[scale=0.50]{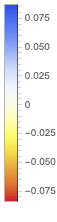}
    \vspace*{0.5cm}\\
    %%%% (21,20) - 1.61 degrees
    (b)
    \includegraphics[scale=0.30]{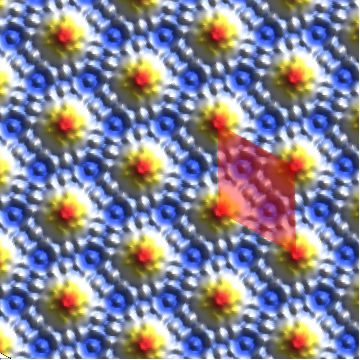}
    \includegraphics[scale=0.50]{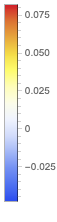}\hspace{0.5cm}
    \includegraphics[scale=0.30]{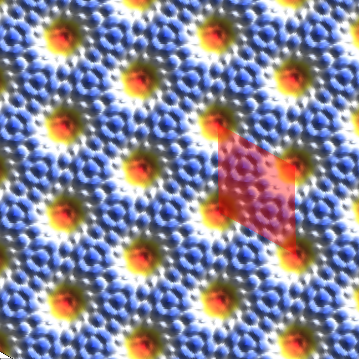}
    \includegraphics[scale=0.50]{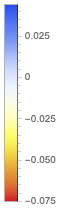}
    \vspace*{0.5cm}\\
    %%%% (13,12) - 2.65 degrees
    (c)
    \includegraphics[scale=0.30]{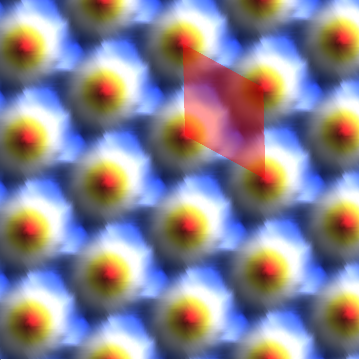}
    \includegraphics[scale=0.50]{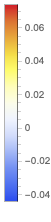}\hspace{0.5cm}
    \includegraphics[scale=0.30]{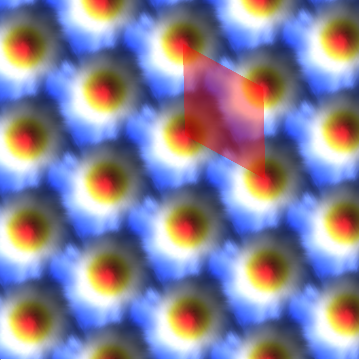}
    \includegraphics[scale=0.50]{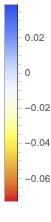}
    \vspace*{0.5cm}\\
    %%%% (9,8) - 3.89 degrees
    \hspace*{0.8cm}
    (d)
    \includegraphics[scale=0.30]{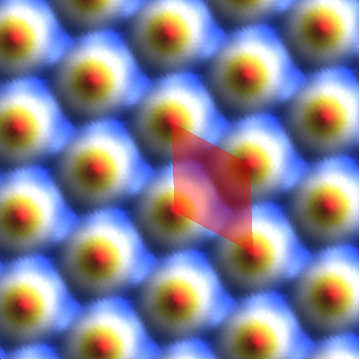}
    \includegraphics[scale=0.50]{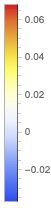}\hspace{0.5cm}
     \includegraphics[scale=0.30]{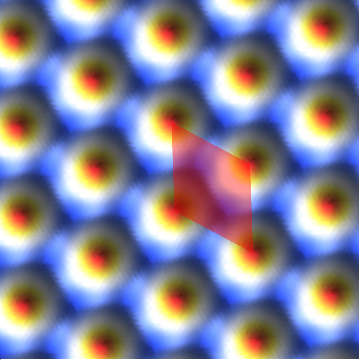}
    \includegraphics[scale=0.50]{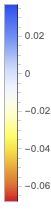}\hspace{1.0cm}
    \caption{The color/relief map of the out-of-plane displacements in TGB at different twist angles: (a) 1.08$^\circ$, (b) 1.61$^\circ$, (c) 2.65$^\circ$, (d) 3.89$^\circ$. In each subset, the left (right) panel refers to the top (bottom) plane. The color bar reports, in each plane, the measure of the $z$-coordinate referred to its mean value in that plane (in {\AA} units). In each map the unit cell is
    highlighted in red.
    }
    \label{fig:out_of_plane}
\end{figure*}

\begin{figure*}
    \includegraphics[scale=0.1]{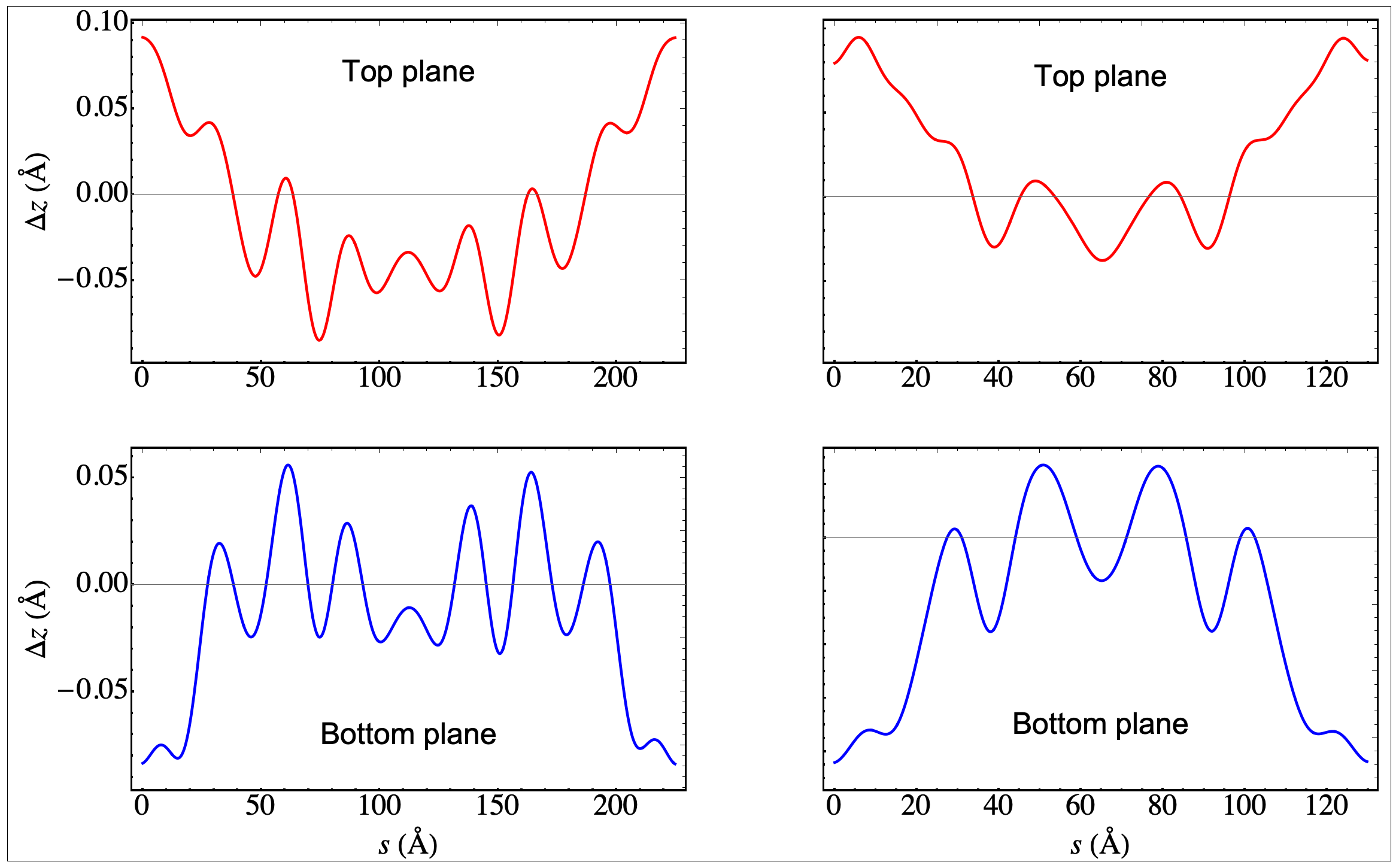}
    \includegraphics[scale=0.1]{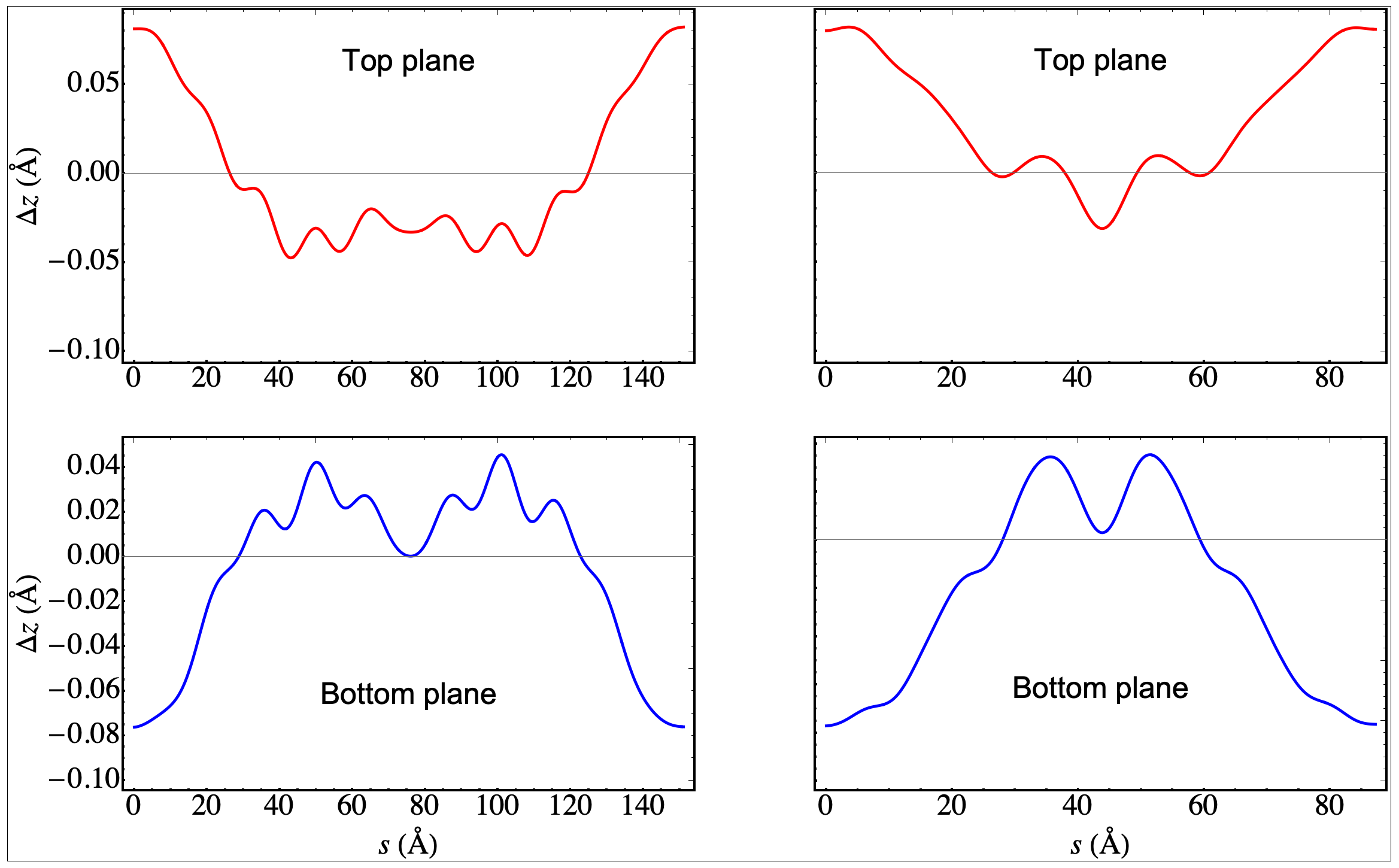}
    \includegraphics[scale=0.1]{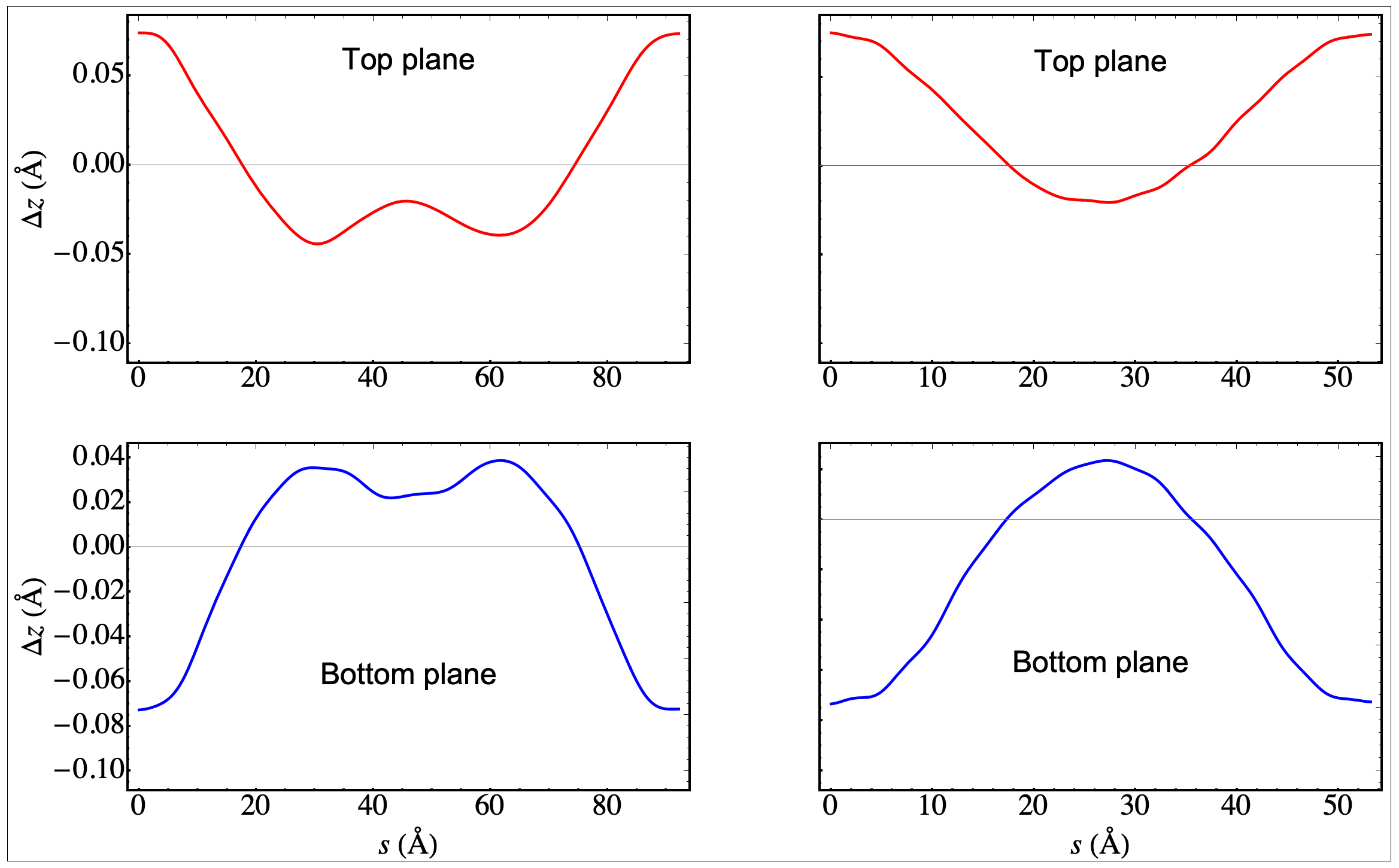}
    \includegraphics[scale=0.1]{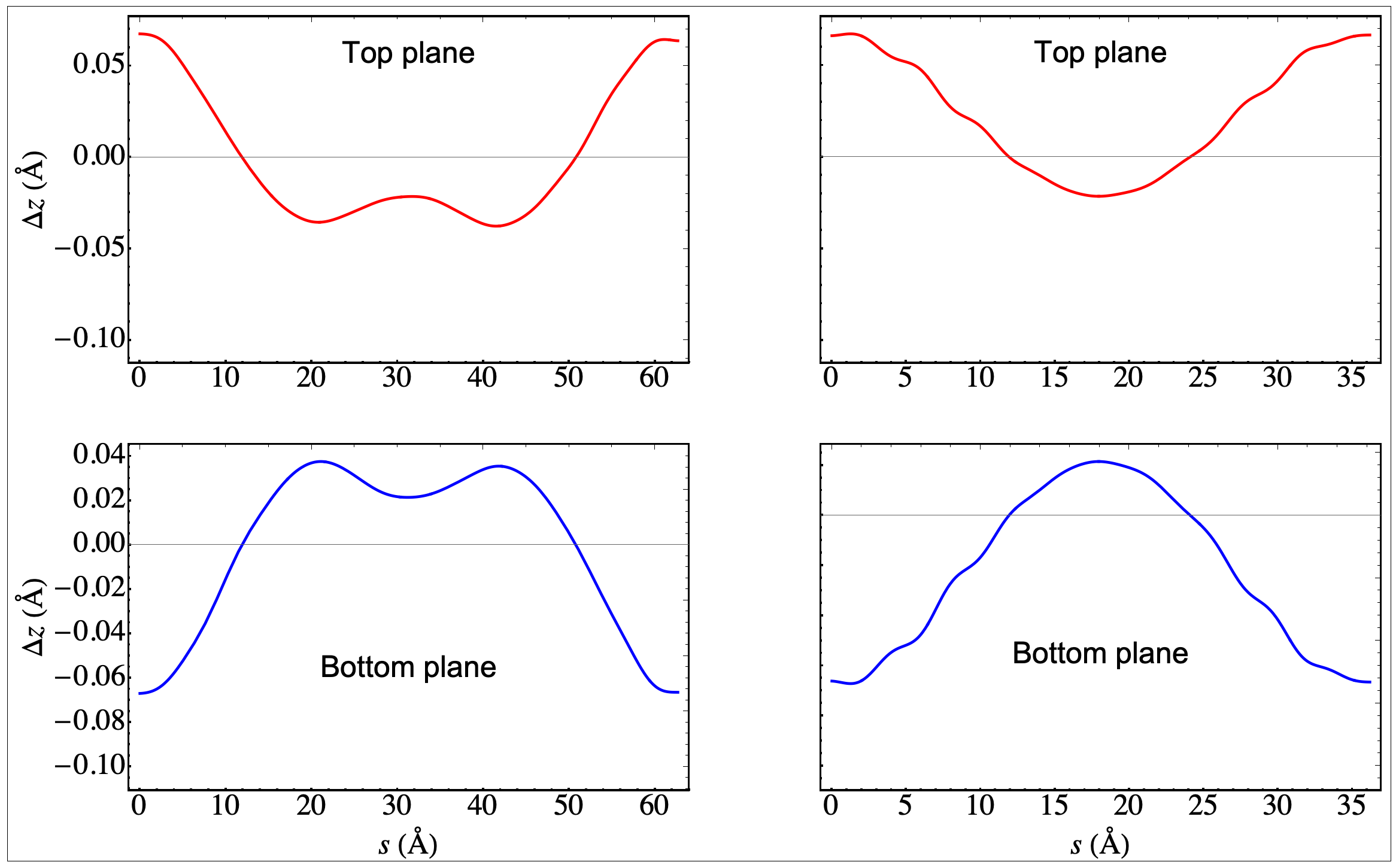}
    \caption{The out-of-plane displacements of each graphene plane along the unit cell long and short diagonal for the four twist angles: $\theta=1.08^\circ$
    (top left box), $\theta=1.61^\circ$
    (top right box), $\theta=2.65^\circ$
    (bottom left box), $\theta=3.698^\circ$
    (bottom right box). 
    The displacements are measured as the difference $\Delta z = z - z_{avg}$ of the $z$
    coordinate of each atom in a given plane and the average $z$ coordinate $z_{avg}$ in that plane
    ($\Delta z=0$ for all atoms in the initial, twisted but unrelaxed configuration). 
    In each box the higher (lower) panels correspond to the top (bottom) plane, whereas the left (right)
    panels correspond to the long (short) diagonal (notice the different scale on the horizontal axis).
    $s$ represents the coordinate along the two diagonals, with $s=0$ corresponding in both cases to the
    unit cell origin (or a lattice equivalent site), where the AA stacking is preserved. To draw this plot the atoms whose projections
    in the $x-y$ plane lie onto are are closest to the unit cell diagonals have been considered.}
    \label{fig:out_of_plane_diagonals}
\end{figure*}

\begin{figure*}
    \centering
     %%%% (31,30) - 1.08 degrees
%    (a)\\
%    \includegraphics[scale=0.34, angle=90]{31_30_top_xy.png}\hspace*{-0.6cm}
%    \includegraphics[scale=0.34, angle=90]{31_30_bottom_xy.png}
%    \includegraphics[scale=0.4]{31_30_bottom_xy_bar.png}
%    %%%% (21,20) - 1.61 degrees
%    (b)\\
%    \includegraphics[scale=0.34, angle=90]{21_20_top_xy.png}\hspace*{-0.6cm}
%    \includegraphics[scale=0.34, angle=90]{21_20_bottom_xy.png}
%    \includegraphics[scale=0.4]{21_20_bottom_xy_bar.png}
%    %%%% (13,12) - 2.65 degrees
%    (c)\\
%    \includegraphics[scale=0.34, angle=90]{13_12_top_xy.png}\hspace*{-0.6cm}
%    \includegraphics[scale=0.34, angle=90]{13_12_bottom_xy.png}
%    \includegraphics[scale=0.3]{13_12_bottom_xy_bar.png}
%    %%%% (9,8) - 3.89 degrees
%    (d)\\
%    \includegraphics[scale=0.34, angle=90]{9_8_top_xy.png}\hspace*{-0.6cm}
%    \includegraphics[scale=0.34, angle=90]{9_8_bottom_xy.png}
%    \includegraphics[scale=0.3]{9_8_bottom_xy_bar.png}
    \includegraphics[scale=0.5]{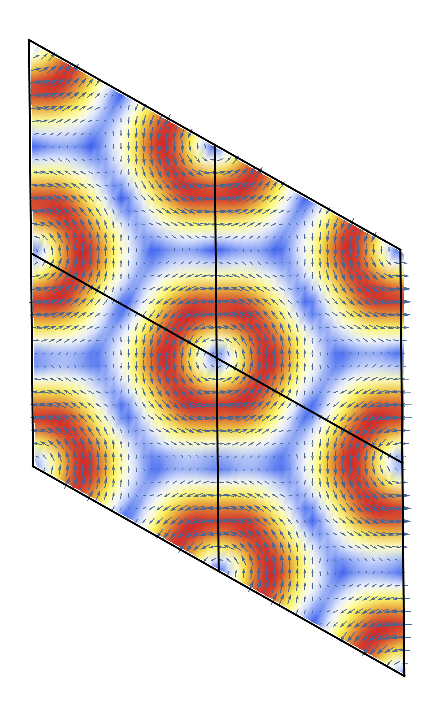}
    \includegraphics[scale=0.5]{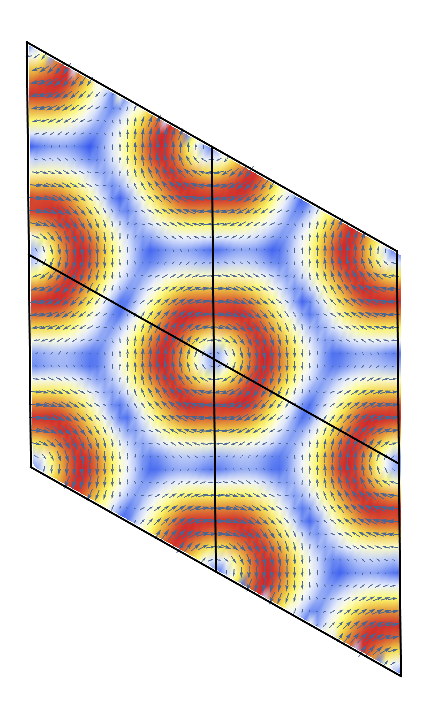}\\
    $\theta=1.08^\circ$
    \includegraphics[scale=0.4]{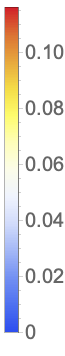}\hspace{1cm}
    $\theta=1.61^\circ$
    \includegraphics[scale=0.4]{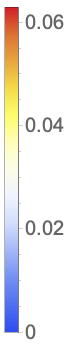}\hspace{1cm}
    $\theta=2.65^\circ$
    \includegraphics[scale=0.4]{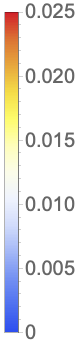}\hspace{1cm}
    $\theta=3.89^\circ$
    \includegraphics[scale=0.4]{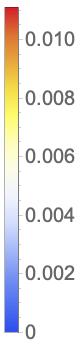}
    \caption{The color map and vector field of the in-plane displacements in TGB for the top (left) and bottom (right) plane. Each vector points in the displacement direction and its magnitude is proportional to the displacement norm. The latter is also highlighted by the color. Since the patterns look quite similar at different twist angles only the result for 1.08$^\circ$ is shown. The color bars report the measure of the displacement in the $x-y$ plane with respect to the unrelaxed structure (in {\AA} units). }
    \label{fig:in-plane}
\end{figure*}

\section{Methods}
\label{sec:methods}
DFT calculations have been carried out using the Vienna Ab initio Simulation Package (VASP) \cite{Kresse1}. 
The vdW-DF2 exchange-correlation functional \cite{Hamada} has been adopted to properly take into account the long-range interactions
taking place between atoms belonging to different graphene layers.
A PAW pseudopotential \cite{Blochl,Kresse2} has been employed for carbon with the 2p orbitals in valence, and the 1s orbitals frozen in the core. The single particle Bloch waves were expanded with a plane wave basis set, using a cutoff energy of 400 eV. 

TBG systems corresponding to four twist angles have been considered: $\theta=1.08^\circ$, $\theta=1.61^\circ$, $\theta=2.65^\circ$, and $\theta=3.698^\circ$. The rotation is carried out starting from two perfectly AA stacked graphene layers and rotating around
an axis orthogonal to the layers and passing through two C atoms, one on top of the other, belonging to the two planes (that therefore preserve their initial AA stacking).
The respective structures can also be classified, according to the notation commonly used in the literature \cite{Shalcross:PRB10}, using the pair
of indexes $(n,m)$: (31,30), (21,20), (13,12) and (9,8), respectively.
The corresponding supercells comprise 11164, 5044, 1876, 868 atoms with lattice parameter $a=$ 129.75, 87.21, 53.19 and 36.18 {\AA}, respectively.

Sampling of the Brillouin zone (BZ) for the self-consistent (SCF) calculations was restricted at the $\Gamma$ point for all four systems,
no significant changes were observed after increasing the size of the sampling of the BZ for the smaller supercells. Single-particle energies at other points in the BZ were obtained by non-SCF calculations. 

For the smallest angle, because of the size  of the simulation cell, we could only compute one $k$-point at a time, and the reported single-particle energies were therefore referred to the Fermi energy computed in the SCF calculation.  The size of the supercell in the direction orthogonal to the layers ($z$-axis) was initially 
fixed at 10 {\AA}, corresponding to about  6.5 {\AA} vacuum space, introduced to prevent periodic replicas of the TBG supercell from interacting with each other. 
Full relaxation of the atomic positions was carried out until the residual forces were smaller than 0.002 eV/{\AA}. Additional calculations were repeated using supercells with $z$-axis of 12 {\AA} and 14 {\AA}. A small residual (maximum) relaxation of less than 0.002{\AA}  was observed as the $z$-axis was increased to 12 {\AA}, but no further relaxation was detectable with the largest 14 {\AA} vacuum space. 
%The initial relaxation was carried out using 2880 physical cores distributed over 80 nodes of a Cray XC-40 machine, over a period of about 30 days. Calculations with 14 {\AA} vacuum required 5760 cores on 160 nodes to accommodate the extra memory requirements. 
All symmetries were turned off.
%For the other angles the vacuum space was set to 10 {\AA} (increased up to 20 {\AA} for the smallest supercell, showing no or negligible
%effect on the calculated electronic and structural properties).
Further  detail on the calculations can be found in our previous paper~ \cite{NoiPRB:2019}.

%\pcom{QUESTI DATI SONO RELATIVI ALLA CELLA 31-30 NON DOVREMMO DIRLO ESPLICITAMENTE ANCHE PER LE ALTRE CELLE O ALMENO DIRE CHE QUESTO E' IL CASO PIU' SFIGATO?}

Tight-binding calculations of the TBG electronic structure at different twist angles and geometries were also carried out using 
the Slater-Koster tight binding parametrization for $p_z$ carbon atoms:
\begin{eqnarray}
\label{hoppingt}
t(R)&=&-V_{pp \pi} \left[1-\left(\frac{\mathbf{R}\cdot\hat {\mathbf{z}}}{R}\right)^2\right]-
V_{pp \sigma} \left(\frac{\mathbf{R}\cdot\hat{\mathbf{z}}}{R}\right)^2\:,\\
V_{pp \pi} &=& V_{pp \pi}^0 e^{(R-a_0)/r_0}, \;\;\;\;\; V_{pp \sigma} = V_{pp \sigma}^0 e^{(R-d_0)/r_0}\:\:.
\nonumber
\end{eqnarray}
Here $r_0=0.184 a$ is the decay length of the transfer integral, $a_0=a/\sqrt{3}$ is the first-neighbor distance in graphene, $d_0=0.335$ nm is the intralayer distance, chosen in agreement with that of graphite. $V_{pp \pi}^0=-2.7 eV$ and  $V_{pp \sigma}^0=0.48 eV$ are the in-plane and-out-of plane nearest-neighbours hopping energy as from Ref.~ \cite{Moon:2013bc}.

We also adopt  a  continuum model generalizing the model proposed in Ref.s~\cite{LopesdosSantos:2007fg,Bistritzer:2011ho,Moon:2013bc,Koshino:2018hm}, providing an effective low-energy band structure. 
Within this approach, 
the two planes are coupled via 
two overlap coefficients $u,u^\prime$, that can be expressed as integrals involving the tight-binding hopping term $t(R)$ of Eq.\ref{hoppingt}. The special case $u=u^\prime$ corresponds to the unrelaxed graphene bilayer.
%Following Ref.\cite{NoiPRB:2019}, 
In order to give a minimal model capable of describing (at least) the low energy properties of the ab-initio band structure, we do not calculate $u,u^\prime$ but use them as fitting parameters. In the following we will show that the fitted parameters are relatively close to (but quantitatively different from) those obtained performing the hopping integrals \cite{NoiPRB:2019}. %of Eq.s~\ref{uup}.
Remarkably, we will show that $u,u^\prime$ can be chosen almost independently of the twist angle.

\section{Geometric relaxation}
\label{sec:geometry}

\begin{figure}
(a) \includegraphics[scale=0.5]{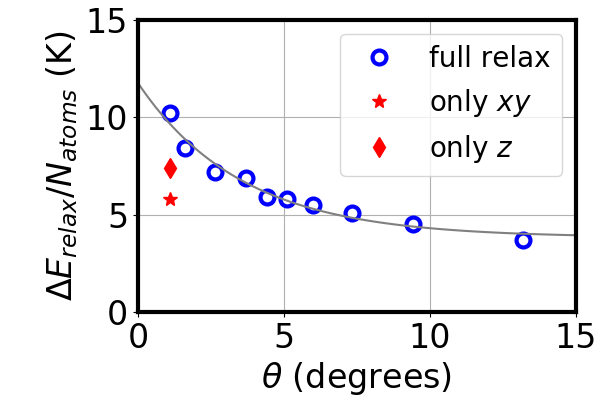}\vspace*{0.8cm}
(b) \includegraphics[scale=0.5]{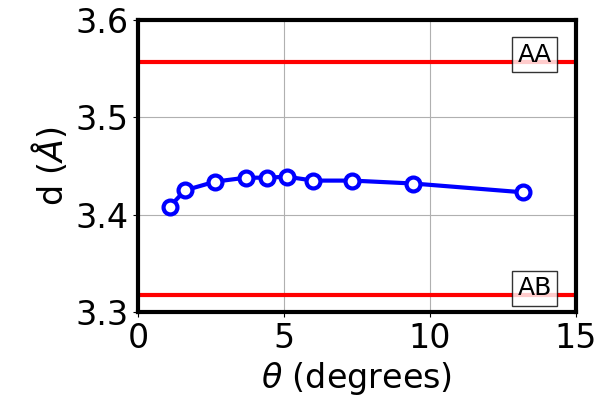}
    \caption{(a) The per-atom relaxation energy as a function of the twist angle (blue circles). The red star (diamond) refers to $\theta=1.08^\circ$ and show the relaxation energy when only the $xy$ ($z$) coordinates are allowed to relax. (b) The interplane distance as a function of the twist angle calculated as the difference between the averages of the $z$ coordinates in each plane. The same distance calculated for the untwisted bilayer, in both the AA and AB stacking is also highligted by the red lines.}
    \label{fig:relax_energy}
\end{figure}

Through a proper inclusion, within the ab initio approach, of the long-range inter-layer vdW interaction, 
we provide a detailed and accurate description of the atomic
relaxations arising from the inter-layer interaction.
We start discussing the out-of-plane atomic displacements. We give two
complementary representations, in Fig. \ref{fig:out_of_plane} and Fig.
\ref{fig:out_of_plane_diagonals}. 
The former shows a color/relief map of the atomic displacements with respect to the ideal, unrelaxed structure.
%The same displacements are represented in the latter figure, but restricted to the atoms distributed along the supercell long and short diagonal.
To properly understand the  results, we should recall that the TBG is built up starting from an ideal AA stacked bilayer, and then
rotating around an axis orthogonal to the graphene planes and passing through
to two atoms, each belonging to a different layer. After twisting, these latter atoms preserve the original AA stacking, that
is kept also after relaxation. However, due to the steric repulsion of
their $p_z$ orbitals, they move far apart from each other. This corresponds to the hills (valleys)
in the top (bottom) plane, clearly visible in the left (right) panels of Fig. \ref{fig:out_of_plane},
highlighted in red. This can also be easier inferred from Fig.
\ref{fig:out_of_plane_diagonals}. Here, for the different twist angles, we show the out-of-plane displacements
of the atoms positioned onto or closest to the unit cell long and short diagonals. Because the starting configuration
is that of two twisted but flat graphene planes, these displacements are visualized,
in each graphene plane, as a difference of the $z$ coordinate of each atom and 
the average $z$ in that plane (highlighted with a thin solid line). We can clearly distinguish how
the two atoms (one for each plane) on the cell corners preserve their initial AA stacking, 
with their final distance estimated to be 3.58 {\AA}. Correspondingly, we can estimate the distance $d$ between the two graphene
planes, as the difference between the averages of the $z$ coordinates in each plane. This is reported in 
Table \ref{table:energies}(a) and shown in Fig. \ref{fig:relax_energy}, where we can observe a reduced
distance at small twist angles. From the figure we can also see
that the interplane distance lies between the calculate interplane distances for the untwisted bilayer at AA and AB stacking (highlighted bi the red solid lines).

The patterns in Fig. \ref{fig:out_of_plane} show that AA stacking regions, where there is an enhanced distance
between an atom in the top plane and the corresponding (closest) atom in the bottom one, alternate with
regions with AB stacking, the latter being predominant. 
One would expect a smooth change of the displacements when moving from AA to AB regions. This naive prediction is verified only for the larger angles. For the smaller twist angles
such smooth behavior is replaced by an oscillating pattern resulting in an atomic corrugation, as clearly visible
from the relief maps in Fig.s \ref{fig:out_of_plane}(a-b).

The out-of-plane displacements are accompanied with significant in-plane relaxations. The (x,y) displacement field is
shown in Fig. \ref{fig:in-plane}. Here the color map and the vector lengths are proportional to the displacement
with respect to the unrelaxed  twisted bilayer. Since the patterns look quite similar at different twist angles only the result 
for $1.08^\circ$ is shown, whereas the color bars distinguish the different systems. Interestingly enough, a vortex-like displacement field shows up in each plane, with the vorticity (intended as the curl of the displacement field) changing sign
when moving from the top to the bottom plane and viceversa. Such result can  be explained by considering that 
AB stacking regions minimize the total energy of the system. Hence,
close to an AA stacking configuration, the atoms of the two layers tend to move in their plane in opposite directions, in order to minimize the overlap of their
orbitals that is maximum at AA stacking. Indeed, it is observed that: i) the in-plane displacement is exactly null for
the AA stacked pair of atoms (no arrow and blue region at the unit cell corners in the figure); ii) the maximum displacements
are observed around the unit cell corners, thus for the atoms that mostly feel the ''repulsion`` due to a stacking that
is quite close to AA; iii) no displacement is observed in the AB stacking regions, as it can be evinced by the blue, hexagonal
regions. While the displacement pattern looks quite similar for all the twist angles (at variance with the out-of-plane
displacements), the magnitude of the displacements decreases by an order of magnitude when moving from the smallest to the
largest twist angle (the maximum displacement being of the order of $\sim$ 0.1 {\AA} and $\sim$ 0.01 {\AA} in the two
cases, respectively).

\begin{table}
\caption{Energy gain induced by the structural relaxation as depicted in Figs. \ref{fig:out_of_plane} and \ref{fig:in-plane} for TBG 
at the four considered twist angles. $N_{\mbox{\tiny atoms}}$ is the number of atoms in each system, $\Delta E_{\mbox{\tiny relax}}$ the difference between the energy of TBG at the optimized geometry and that of the same bilayer in the initial configuration (twisted
but unrelaxed graphene bilayer). The optimized distance $d$ between the two graphene planes, computed as the difference between the averages of the $z$ coordinates in the two planes, is also reported. }
\begin{ruledtabular}
\begin{tabular}{ l c c c c }
                                                 & \bf{(9,8)}  & \bf{(13,12)} & \bf{(21,20)}  & \bf{(31,30)} \\
$N_{\mbox{\tiny atoms}}$                         & 868    & 1876    & 5044     & 11164 \\                   
$\Delta E_{\mbox{\tiny relax}}$ (meV)                  & -0.513 & -1.157  & -3.657   & -9.797\\
$\Delta E_{\mbox{\tiny relax}}/N_{\mbox{\tiny atoms}}$ (meV) & -0.590 & -0.617  & -0.725   & -0.878 \\
$\Delta E_{\mbox{\tiny relax}}/N_{\mbox{\tiny atoms}}$ (K)   & 6.9    & 7.2     & 8.4      & 10.2 \\
$d$ ({\AA})                                                 & 3.438  & 3.434   & 3.425    & 3.408
\end{tabular}
\label{table:energies}
\end{ruledtabular}
\end{table}

The relaxation energy,  is the energy gained by the structure when it is allowed to relax with respect to the unrelaxed configuration, in which the atoms are arranged in  ideal honeycomb lattices in two parallel planes. It is reported in Table \ref{table:energies}. Such gain is extensive, i.e. it depends on the number of atoms in the unitary cells. In order to allow for a fair comparison between different twist angles, in Table \ref{table:energies} we also report the energy gain per-atom. The relaxation energy normalized to the number of atoms in the unit cell, is plotted  in Fig. \ref{fig:relax_energy}(a) as a function of the twist angle (blue dots). In addition to the four angles in Tab.\ref{table:energies}, we show also points relative to the angles $4.41^\circ,5.09^\circ,6.01^\circ,7.34^\circ,9.43^\circ,13.17^\circ$ (corresponding to $(n,m)=(8,7),(7,6),(6,5),(5,4),(4,3),(3,2)$ respectively), to better represent the limit of large twist angles (small unit cells). 
%The grey line is an exponential fit of the data
%, corresponding  to 

%\begin{equation*}
%     \frac { \Delta E_{ %\mbox{\tiny relax} } } { N_{ %\mbox{\tiny atoms} } } = 3.824 %+ 7.938
%     \exp \left( -0.281 \theta %\right)
%\end{equation*}
%where the energies are in K. 
%\pcom{CHE NE PENSATE DI QUESTO FIT ESPONENZIALE? UN'IDEA PERCHE' LA DIPENDENZA FUNZIONALE POSSA ESSERE QUESTA?   }
The per-atom relaxation energy increases from 3.7 K to 10.2 K when passing from the largest twist angle (corresponding to the smallest unit cell) to the smallest twist angle (corresponding to the largest unit cell).  Such relaxation energy is an estimate of the upper bound of the thermal energy that in a real experiment would 
induce thermal fluctuations of the atomic positions that in turn would destroy   the ground-state geometry pattern shown in Figs. \ref{fig:out_of_plane} and \ref{fig:in-plane}. In the case of the $1.08^\circ$ magic angle it has to be compared, for instance,  with the critical temperature at which
zero-resistance states are observed ($T_c \le 1.7 K$~\cite{Cao:2018ff}) or at which the correlated insulator behavior at half-filling 
is experimentally observed ($T < 1.0 K$~\cite{Cao:2018kn}).
Interestingly, the relaxation energy increases upon decreasing the twist angle. This is explained by considering that in the case of large unit cells the atoms have more freedom to relax to lower total energy.
In Fig. \ref{fig:relax_energy} b) we also show the interplane distance (averaged over the unit cell). It is almost half way between the distance of the AA and the AB bilayer. However, as we have aready noticed before, relaxation mechanisms tend to enlarge the "effectively" AB regions, hence the average distance is slighly closer to that of the AB stacking. 

\subsection{Focus on the first magic angle at $\theta=1.08^\circ$}We now focus on TBG at the first magic angle, $\theta=1.08^\circ$,
which shows the most intriguing and marked relaxation pattern. In the following, we investigate  the interplay between the
out-of-plane and in-plane relaxations.  
This is done performing two optimizations, the first by allowing all atoms to relax only along the  $z$ direction, the second by allowing them to relax only their graphene planes (hence the $z$ coordinates are fixed while $xy$ coordinate are free to move).  
These partial relaxations have nontrivial effects on the electronic properties of the TBG. 

First of all, 
we notice that these partial relaxations lead to a smaller gain  in the total energy with respect to a full relaxation. 
This is shown
by the red star and diamond in Fig. \ref{fig:relax_energy}.
This result is not surprising and is due  to the reduced freedom of the atoms (constrained along $z$ or $xy$) to find optimal minimal energy configurations.
Hence, neither the only-$xy$, nor the only-$z$ relaxations catch most of the relaxation energy (which for this system amount to 
$\sim$ 10 K), that is due to an interplay between them. Such an interplay will result even more evident in the electronic properties, 
when we will discuss the energy gaps separating the FB manifold from the lower and higher energy bands.

\section{Electronic Properties}
\label{sec:electron}
\begin{figure*}
    \includegraphics[scale=0.4]{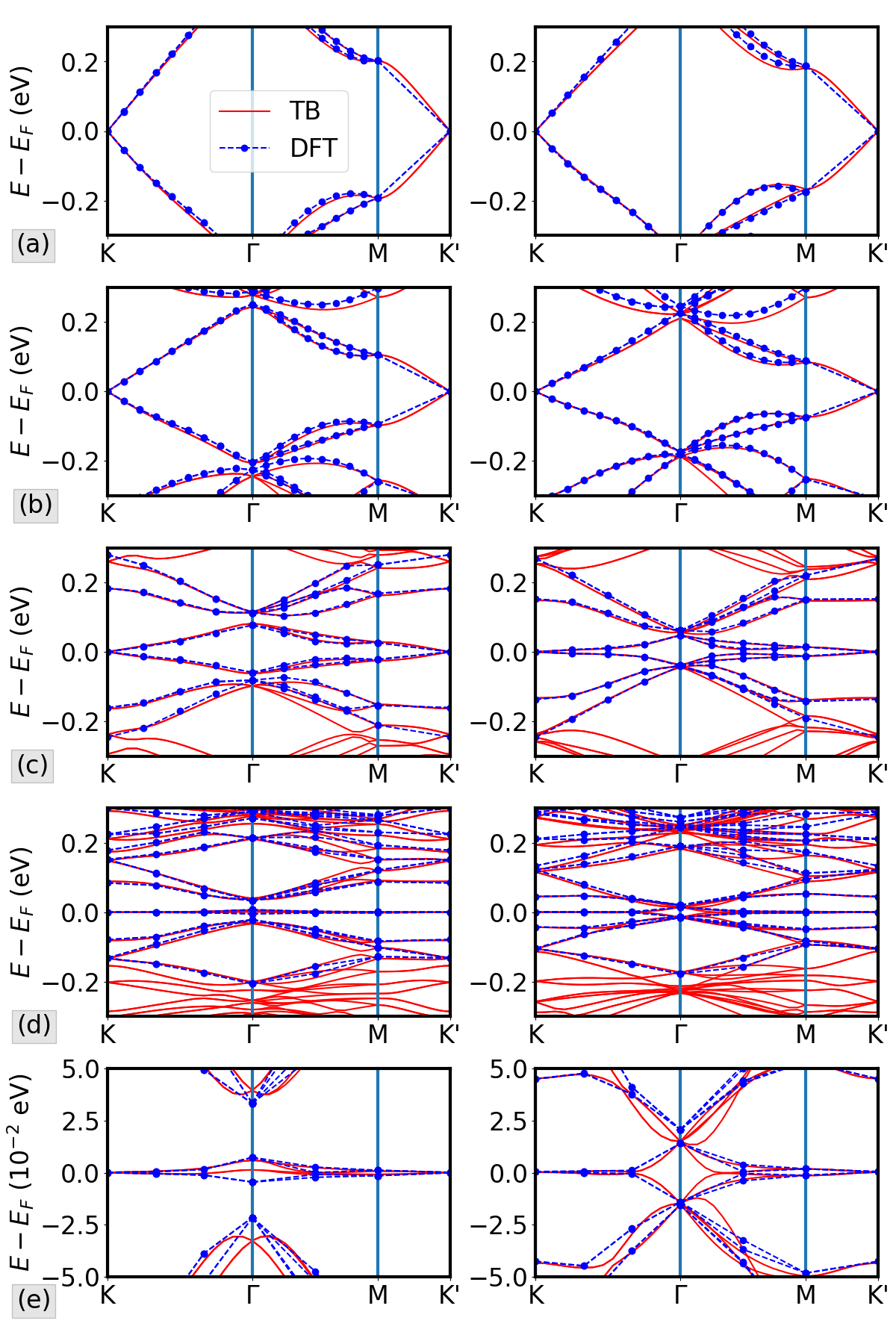}
    \caption{Electronic band structure of TBG along the $K-\Gamma-M-K'$ path at different twist angles: 
    (a) $\theta=3.698^\circ$, (b)  $\theta=2.65^\circ$, (c) $\theta=1.61^\circ$, (d) $\theta=1.08^\circ$,
    (e) $\theta=1.08^\circ$ zoomed around the Fermi energy (notice the different scale on the $y$ axis). The left (right)
    panels corresponds to the relaxed (initial) structure.
    Blue dashed lines and filled dots correspond to the DFT calculation, red solid lines to the TB approach.
    Zero energy corresponds to the Fermi energy. The effective continuum model mentioned in Sec. \ref{sec:methods} can be also used to approximate the DFT calculations. The parameters $u, u^\prime$ are independent of the angle and, within a reasonable error, can be approximated as $u=u^\prime=0.107\pm0.004$ in the unrelaxed case and $u=0.078\pm0.002$, $u^\prime=0.098\pm0.004$ for relaxed structures.}
    \label{fig:bands}
\end{figure*}
In this section we discuss the electronic structure for the four angles analyzed, with and without including relaxation mechanisms.  
The results are summarized in Fig.s \ref{fig:bands}, in which we plot the band structure at the angles $\theta=3.698^\circ$ (panels a), $\theta=2.65^\circ$ (panels b), $\theta=1.61^\circ$ (panels c), $\theta=1.08^\circ$ (panels d-e). The left panels show the band structures including relaxation mechanisms, while in the right panels the atoms are fixed in their lattice positions. In
Fig.~\ref{fig:bands}(e) we show a zoom close to the Fermi energy in the case of the magic angle $\theta=1.08^\circ$. The blue points correspond to the bands calculated using the DFT approach described in Sec. \ref{sec:methods}. 
The red (full) lines are obtained  within a tight binding calculation where, however, the atomic positions for the relaxed structures 
are the ones optimized within the DFT approach. In general, we notice that there is an excellent agreement between the 
red (full) lines and the blue points. The DFT calculation could be well approximated also adopting a continuum model 
(see Sec. \ref{sec:methods}), where the interplane hoppings are parametrized, for the optimized structure, by coefficients $u=0.078\pm0.002$, $u^\prime=0.098\pm0.004$ and for unrelaxed ones by $u=u'=0.107\pm0.004$ almost independently of the angle. 
That can be adopted as a minimal single particle description within more complex many body \cite{Koshino:2018hm,Xie:2019aa} approaches in order to study the superconductivity and the Mott insulating state of this system.

Inspection of Fig. \ref{fig:bands} shows that, comparing the TB and the DFT curves,
there is a tiny mismatch that can be evidenced, for instance, at $\Gamma$. This is due to the fact that, despite sharing the same atomic position, the two approaches do not account for e-e interaction in the same way. In particular, the tight binding approach does not
takes into account any e-e interaction while in the DFT approximation the e-e is accounted within the local density approximation. This discrepancy, at the $\Gamma$ point, could indeed be an estimate of the Hartree energy in TBG.

In all panels, it is clearly visible that relaxation mechanisms tend to maximize the energy gaps at $\Gamma$ and, in particular, in the case of the magic angle $\theta=1.08^\circ$ the gap separating the FBs from the higher (lower) energy bands is about
26 meV (16 meV), consistent with the experiments\cite{Cao:2018kn,Cao:2018ff,Kerelsky:2019aa,Xie:2019aa,Jiang:2019aa,Choi:2019aa,Lisi2020direct}. On the other hand,
those gap cannot be reproduced at all if no relaxation is allowed. Aimed at identifying if there is a predominant role
of the in-plane or of the out-of-plane displacements, these energy gaps have been estimated also when the system is allowed to
relax only either in the $x-y$ plane or along the $z$ axis: in the former case, we obtained gaps of  $\sim 2$ 
and $\sim 14$ meV, that underestimate these values in the fully relaxed system, especially in the hole side; in the latter case,
the two gaps turn out to be both zero, showing that a quite important role is played by the $x-y$ displacements.

As far as the FB dispersion is concerned, we obtain a full bandwidth of $\sim 20$ meV in the relaxed system, 
which is of the order of the one measured in the experiment of Ref.~\onlinecite{Cao:2018ff}.%, to be contrasted with the $\sim 12$ eV bandwidth measured in the unrelaxed system.

%First of all, we may notice that the FB has a dispersion of
%∼20 meV (calculated ab initio), which is almost twice the one measured in the experiment of Ref. [2]. Another relevant issue is that the 
%unrelaxed ab initio calculation is not able to reproduce the gap between the FB and the first excited bands (both on the electron side and %on the hole side) that are responsible for the band insulating phases.
%The FB now extends for ∼12 meV around the Fermi level this number is subject to an error of approximately 3 meV). That bandwidth is in 
%good agreement with the one measured in experiments (��10 meV; see Ref. [1]). It is separated by a gap of 26 meV (16 meV) from the highest %occupied (low- est unoccupied) bands, which should be compared with the thermal activation gap of ∼40 meV measured in experiments. Such
%discrepancy is not much larger than the convergence error in the DFT calculations. 
%\pcom{qui osservazioni sulle gap coi rilassamenti parziali che lascio a te}
%Gaps at gamma
%only xy 0.002200  0.014400,
%only z 0 0
%both  0.026  0.016

\section{Conclusions}
\label{sec:conclusions}
The FBs in TBG at the first magic angle can be intimately related with the atomic displacements arising as effect of the
interlayer vdW interaction. Large scale first principles calculations allow us to conclude that  experimental gaps cannot
be reproduced if we consider a flat bilayer system. Relaxation effects are thus crucial as already noticed in our previous manuscript \cite{NoiPRB:2019}.   In this manuscript, we have also investigated  partial relaxation processes with only in-plane or   out-of-plane displacements, however the resulting band structure does not show the expected gaps in both cases.
Out-of-plane relaxations are characterized by a strongly oscillating pattern, that is smoothed (until it disappears) at large
twist angles.
On the other hand, the in-plane displacements show a vortex-like configuration, where the vorticity assumes opposite values in the two planes. However, the magnitude of these displacements decreases upon increasing the twist angle.
Overall, the combination of the two patterns allow the atoms to override the steric repulsion felt by the $p_z$ orbitals, maximizing the regions with AB stacking  at expenses of the regions showing AA stacking.

The energy gain induced by the relaxation is the larger, the smaller is the twist angle, decreases by increasing the twist angle and eventually seems to reach a plateau of $\sim$6.8 K at large twist angles. The smaller angles
correspond to larger  unit cells, that can easier accommodate
the atomic rearrangement, and  correspond to large energy gain,
of the order of $\sim$10 K, much larger than the temperature at which the most exotic phenomena,
such as correlated insulating phase and superconductivity are detected. Since the latter are intimately
related to the presence of the FBs, that in turn we demonstrate being related with the atomic relaxation, such temperature (10 K) should be considered as an upper limit. Higher temperatures would destroy relaxation effects
due to thermal atomic oscillations, and as a consequence FBs effects would be hindered. 

Including relaxation effects not only we reproduce band gaps consistent with the ones measured in experiments, but we also give the effective parameters of a low energy continuum model to be adopted for further investigation including correlation effects. Interestingly enough, despite the fact that relaxation patterns have different shapes at various angles, the interplane hopping coefficients $u,u'$ are found to be almost independent of the twist angle.  That could be a useful hint to apply the continuum model also to smaller angles, where the unit cell would become umpractically large to be attacked with atomistic  approaches.

% If you have acknowledgments, this puts in the proper section head.
\begin{acknowledgments}
% put your acknowledgments here.
G.C. and P.L. contributed equally to this work. We thank A. Stroppa for stimulating discussions. We acknowledge use of the Monsoon2 system, a collaborative facility supplied under the Joint Weather and Climate Research Program, a strategic partnership between the UK Met Office and the Natural Environment Research Council.
\end{acknowledgments}

% Create the reference section using BibTeX:
%\bibliography{TBG}
%apsrev4-2.bst 2019-01-14 (MD) hand-edited version of apsrev4-1.bst
%Control: key (0)
%Control: author (72) initials jnrlst
%Control: editor formatted (1) identically to author
%Control: production of article title (-1) disabled
%Control: page (0) single
%Control: year (1) truncated
%Control: production of eprint (0) enabled
%

\end{document}